# Supernova progenitor stars in the initial range of 23 to 33 solar masses and their relation with the SNR Cas A.

B. Pérez-Rendón[1,2], G. García-Segura[2], and N. Langer[3]

[1] Departamento de Investigación en Física. Universidad de Sonora. Apdo. Postal 5-088. Hermosillo, Sonora, México.
[2] Instituto de Astronomía, Universidad Nacional Autónoma de México, Apdo. Postal 877, Ensenada 22800, Baja Californa, México.
[3] Astronomical Institute, Utrecht University, P.O. Box 80000, 3508 TA Utrecht, The Netherlands.

**ABSTRACT**

*Context.*
*Aims.* Multi wavelength observations of Cassiopeia A (Cas A) have provided us with strong evidence of circumstellar material surrounding the progenitor star. It has been suggested that its progenitor was a massive star with strong mass loss. But, despite the large amount of observational data from optical, IR, radio, and x-ray observations, the identity of Cas A progenitor is still elusive.
*Methods.* In this work, we computed stellar and circumstellar numerical models to look for the progenitor of Cas A. The models are compared with the observational constraints that come from chemical observed abundances and dynamical information.
*Results.* We first computed stellar evolution models to get time-dependent wind parameters and surface abundances using the code STERN. To explore the range of masses proposed by several previous works, we chose a set of probable progenitor stars with initial masses of 23, 28, 29, 30, and 33 $M_\odot$ , with initial solar composition (Y=0.28, Z=0.02) and mass loss. The derived mass loss rates and wind terminal velocities are used as inner boundary conditions in the explicit, hydrodynamical code ZEUS-3D to simulate the evolution of the circumstellar medium. We simplified the calculations by using one-dimensional grids in the main sequence and red super giant (RSG) stages, and two-dimensional grids for the post-RSG evolution and supernova (SN) blast wave.
*Conclusions.* Our stellar set gives distinct SN progenitors: RSG, luminous blue super giants (LBSGs), and Wolf-Rayet (WR) stars. We named these type of stars "luminous blue super giant" (LBSGs) to distinguish them from normal blue super giants (BSGs) of much lower initial masses. The 23 $M_\odot$ star explodes as an RSG in a $\rho \sim r^{-2}$ dense, free-streaming wind surrounded by a thin, compressed, RSG shell. The 28 $M_\odot$ star explodes as an LBSG, and the SN blast wave interacts with a low density, free streaming wind surrounded by an unstable and massive "RSG+LBSG" shell. Finally, the 30 and 33 $M_\odot$ stars explode as WR stars surrounded by fast WR winds that terminate in a highly fragmented "WR+RSG shell". We compared the surface chemical abundances of our stellar models with the observational abundances in Cas A. The abundance analysis shows that the progenitor was a star with an initial mass of about 30 $M_\odot$ , while the hydrodynamical analysis favors progenitors with initial masses around 23.

## 1. Introduction

Cas A is the youngest galactic SN remnant (~ 327 years old Thorstensen et al. 2001) and it has been extensively studied in all wavelengths from radio to $\gamma$-rays. The observational data obtained in all wavelengths has provided us with a wide set of observational constraints that, taken together, put strong limits on the nature of its progenitor. Optical emission in Cas A shows an expanding ring of ~ 4' in diameter, which gives a radius of 2 pc, assuming a distance of 3.4 Kpc (Reed et al. 1995), and it comes mainly from three distinctive types of objects: (1) The fast moving knots (FMKs) with high velocities (~ 4,000 - 6,000 km s$^{-1}$ ) that seem devoid of hydrogen and are enhanced in oxygen and other heavy elements synthesized from stellar O burning (Si, S, Ca, Ar). Their high velocities and chemical composition imply that they are material ejected in the SN explosion as shrapnel from the stellar core or mantle (Johnston & Yahil 1984). Most of the FMKs with ~ 6,000 km s$^{-1}$ are located in the optical main shell at ~ 2 pc, although faster FMKs lies on the "jet". (2) The quasi stationary floculii (QSFs) are optical knots with lower velocities (< 400 km s$^{-1}$ ), strong spectral lines of helium and nitrogen, and hydrogen deficient. QSFs show high N/H and He/H ratios (N/H ~ 10 and He/H ~ 5 - 10; Chevalier and Kirshner 1978) and they are located closer to the star, with average radii of ~ 1.5 pc. (3) The nitrogen knots (NKs) have the highest velocities (~ 7,000 - 9,000 km s$^{-1}$ ) and are distributed outside of the main, optical emission shell at an average radius of ~ 3 pc. Detected by Fesen (2001), the NKs spectra are dominated by two [N II] lines in the 4,500 - 7,500 A spectral range, and apparently, the NKs were ejected with a uniform, high velocity. Their N to H$\alpha$ flux ratios imply N/H ~ 10 - 30 times solar abundances.

Cas A also shows strong emission in other wavelengths: radio emission is generally associated with the contact discontinuity (Vink 2005). Bulk x-ray emission comes from hot material shocked by the SN shock fronts (Fabian et al. 1980). Soft x-rays have been detected associated to QSFs (De Laney et al. 2004) and hard x-ray filaments are associated with the forward and reverse shocks. The external x-ray filaments are associated with the forward shock wave at an average distance of 153" ± 12" (2.5 ± 0.2 pc) to the emission center, and evidence has been found about the reverse shock position in an average radius of 95" ± 10" (1.6 ± 0.17 pc) to the x-ray emission centroid (Gotthelf et al. 2001). Proper motion measurements made in optical, radio, and x-rays show variations in the expansion rate from 0.1 % to 0.3 % yr$^{-1}$ (Thorstensen et al. 2001).

Despite all the observational data from optical, infrared, radio, and x-rays, the identity of Cas A progenitor has remained ambiguous. It has been suggested that the progenitor was a massive star with a strong mass-loss rate. Using optical abundances, Peimbert and van den Bergh (1971) and Chevalier (1976) proposed that the Cas A progenitor star lost the H-He envelope previous to the SN explosion and that the QSFs are remnants of that material. Lamb (1978) explained the nitrogen enrichment of QSFs as the result of CNO burning in combination with mass loss. Based on the nitrogen abundances, many authors have proposed that Cas A is the result of the explosion of a WN star (Peimbert 1971; Fesen et.al. 1987). QSFs are thought to be RSG wind-shocked by the WR progenitor wind previous to the SN (García-Segura et al. 1996). The NKs are thought to be expelled from the photosphere of the progenitor star at the SN explosion, and they are the evidence that the progenitor star must have a nitrogen-rich photosphere, but is H-deficient. Some authors suggested a RSG progenitor (Chevalier & Oishi 2003) or a blue super giant (BSG) progenitor (Borkowski et al. 1996), based on the CSM and shock front dynamics. Binary scenarios for the Cas A progenitor have also been discussed by Young et al. (2005). However, most authors agree that the progenitor stellar mass (at ZAMS) lies between 20 and 35 $M_\odot$ with strong mass loss (Fabian et al. 1980; Fesen, Becker & Blair 1987; Jansen et al. 1988; Vink et al. 1996; García-Segura et al. 1996, Young et al. 2005).

Generally, hydrodynamical calculations of the CSM assumed idealized winds with constant properties, but the stellar winds depend on stellar luminosities, masses, radii, and metallicities so the wind parameters change as the stars evolve (Nieuwenhuijzen & de Jager 1990). More realistic models, however, require a prescription for the mass-loss rate and stellar wind velocity as the star evolves in the Hertzpung-Rusell (HR) diagram, coupling the evolution of the CSM to the star itself. The stellar wind interacts with the surrounding matter and the behavior of the CSM is tightly bound to the stellar evolution.

In this work we have combined the observed chemical abundances and dynamical information to make constraints for the Cas A progenitor, and compare the observations with our models to find the progenitor mass. We computed theoretical models of stellar and circumstellar evolution for stars in the mass range generally accepted by most authors.

We used the hydrodynamic, stellar evolution code STERN to build up a set of evolutionary, non-rotating, stellar models with initial masses in the range from 23 to 33 $M_\odot$, with solar composition and mass loss. From our stellar models, we obtained surface chemical abundances, mass-loss rates, and wind velocities as a function of time during the whole stellar evolution, from zero age main sequence (ZAMS) up to end-of-core O burning (i.e., almost pre-SN stage). Thereafter, we used the mass-loss rates and wind velocities, and a prescription for the SN ejecta, to build up numerical models of the CSM evolution, to compare them with the dynamical and morphological features of Cas A. To simulate the CSM and the SN blast wave, we used the ZEUS-3D code, using the stellar evolution outputs as inner boundary conditions.

The paper is organized as follows. Section §2 describes the computational methods and the input physics used in our numerical models. Section §3 describes the stellar evolution models from ZAMS to pre-SN. Section §4 shows the hydrodynamical simulations of the CSM covering the whole stellar evolution. The hydrodynamical simulations also includes the SN ejecta and the interaction with the CSM for 1,000 yr after the explosion. Section §5 compare our numerical results with several observational constraints for Cas A. Finally, §6 gives a summary and our conclusions.

## 2. Computational methods

The numerical methods used in this work are similar to those in García-Segura et al. (1996, hereafter GLM96AB). The stellar models were produced with the stellar code STERN (Langer et al. 1985, Heger 1998) for stars with 23, 28, 29, 30 y 33 $M_\odot$ (at ZAMS) with solar metallicity (Y = 0.28; Z = 0.02) and meteoritic proportion by Grevesse and Noels (1993). All models assume single stars without stellar rotation. We adopted the OPAL opacities by Iglesias and Rogers (1996), equation of state from Langer et. al. (1988), and an external boundary condition to set a stellar atmosphere in expansion (Heger & Langer 1996). The network of nuclear reactions includes 35 isotopes, $^0$n, $^{1-2}$H, $^{3-4}$He, $^{6-7}$Li, $^{7-9}$Be, $^{8-10-11}$B, $^{11-12-13}$C, $^{12-14-15}$N, $^{16-17-18}$O, $^{19}$F, $^{20-21-22}$Ne, $^{23}$Na, $^{24-25-26}$Mg, $^{26-27}$Al, $^{28-29-30}$Si, $^{56}$Fe, and the reaction rates from Caughlan & Fowler (1988), except for the $^{12}C(\alpha,\gamma)^{16}O$ rate where we used the rate of Caughlan & Fowler (1988) multiplied by 1.7. We used the Ledoux criterion for stellar convection, and the processes of convective transport was considered as diffusive processes according to Langer et al. (1985). Convective overshooting was taken into account during core H-burning with an effective convective, overshooting distance of 0.2 times the local pressure scale height (Stothers & Chin 1991). Also we adopt a mixing-length parameter of $\alpha = L/H_P = 1.0$ in the main sequence. In the RSG the $\alpha$ parameter was set to 2.0 to avoid dynamical instabilities in the outer envelope of our models related to high radiation pressure and partial ionization of the gases in the outer layers (Stothers & Chin 2001). In the calculations, we adopted the empirical formulation of the mass-loss rate by Nieuwenhuijzen & de Jager (1990) for all evolutionary stages, except in the RSG and WR phases. Mass-loss rates in the upper end of the HR diagram are uncertain by a large factor, particularly for RSG stars (Chiosi & Maeder 1986). In the RSG phase we used the Nieuwenhuijzen & de Jager mass loss rate multiplied by a factor of 2.0, to allow post-RSG evolution in our models with masses higher than 30 $M_\odot$, according to empirical observations by Humphreys & Nichols (1985). If the surface hydrogen mass fraction fall below $X_s < 0.4$ the star begins a WR phase, and the mass-loss rate was chosen according to Langer (1989). More recent WR mass-loss rates, slightly lower than ours, have been determined by Nugis & Lamers (2000), so that, our WR mass loss rates must be considered as an upper limit. The WN-H rich (or WNL) phase starts when a star becomes a WN star (surface-H < 0.3) while WN-He (WNE) phase begins when the surface hydrogen becomes zero; WC/WO indicate the stage when surface carbon abundance increases $C_s > 0.03$.

The stellar evolution models was computed from ZAMS up to O core exhaustion previous to Si burning. At this stage, the star is in the pre-SN stage, a few days or hours from the SN explosion.

The derived mass-loss rates and terminal wind velocities are used as time-dependent, inner boundary conditions in the magnetohydrodynamical code ZEUS-3D, a three-dimensional, finite-difference, Eulerian explicit code (Stone & Norman 1992, Clarke 1996). This code integrates the hydrodynamical ideal gas equations in the absence of viscosity. We used spherical coordinates, with periodic boundary conditions in the alt-azimuthal direction. The simulations we are done in one-dimensional (1D) or two-dimensional (2D) grids as a function of the stellar evolutionary stage. We calculated 1D simulations in the main sequence and red super giant phases because of the dynamical behavior of their swept-up shells (GLM96AB). Post-RSG evolution was computed on 2D grids because of the strong unstable behavior of the formed structures, especially in WR phases (GLM96AB). The size of the computational 1D domain is $r_{max}$ = 50pc, with 1,000 radial zones, giving a spatial resolution of 0.05 pc/cell. Hereafter, the variables were interpolated on smaller 2D grids to compare with the real dimensions of Cas A. The 2D grids have a computational domain of r = 15 pc × $\phi$ = 30°, with 500 × 300 cells giving a resolution of 0.03pc ×0.1°.

## 3. Stellar evolution models

### 3.1. 23 M$_\odot$ model

Figure 1 shows the path followed by the 23 M$_\odot$ model in the HR diagram, and the evolutionary sequence is shown in Table 1. The H-core burning phase lasts 6.374 Myr, and the star loses 0.81 M$_\odot$ as a fast wind (V$_\infty$ ~ 1,000 - 3,000 km s$^{-1}$) with an average mass-loss rate of 1.3 ×10$^{-7}$ M$_\odot$ yr$^{-1}$. After H-core burning exhaustion the star ignites H shell burning, and late He core burning at 6.4 M yr, becoming an RSG. The cusp in the HR diagram is related to the change in the mixing length parameter.

In Figure 2 the surface chemical evolution of the 23 M$_\odot$ star is shown. The surface chemical composition does not change during MS, until a strong dredge-up at the RSG phase transports H-burning CNO products to the surface: H, O, and C diminish while He and N increase. The RSG envelope abundances become homogeneous because of the convective efficiency. In the RSG, the mass loss rate increases to 1.7 ×10$^{-5}$ M$_\odot$ yr$^{-1}$ and the star loses 9 M$_\odot$ in a dense and slow wind enriched with He and N from the CNO processing.

The RSG mass loss of the 23 M$_\odot$ star is not large enough to produce a blueward evolution in the HR diagram, and the star, after core He-burning, ignites C, Ne, and O-core burning in the RSG phase (Figure 1). The resulted pre-SN model is a 13 M$_\odot$ RSG, with a H-rich envelope of 6.8 M$_\odot$ and a stellar radius of R ~ 700 R$_\odot$, T$_{eff}$ ≈ 4,000 K, and log L/L$_\odot$ ≈ 5.1. According to Heger et al. (2003), a pre-SN star with an envelope mass higher than 3M$_\odot$ would produce an SN IIp.

### 3.2. 28 M$_\odot$ model

This model evolves according to the evolutionary channel shown in Table 1 and evolutionary path in Figure 1. The star loses 1.39 M$_\odot$ in the MS during 5.252 Myr with a mass-loss rate of 2.8 × 10$^{-7}$ M$_\odot$ yr$^{-1}$. After core H-burning exhaustion the star becomes an RSG during 3.3 ×10$^5$ yr. In the RSG stage, the stellar hydrogen envelope becomes enriched with He and N from the CNO, H-shell burning. During the RSG phase, the 28 M$_\odot$ star loses 16.9 M$_\odot$ and evolves to the blue side of the HR diagram, performing a blue loop. During the blue loop, we have a new increase in He and N surface abundances (Figure 2), whereas the central He mass fraction drops from $Y_c$ = 0.469 (blue loop onset) to $Y_c$ = 0.218 at the blue tip (t = 5.73 Myrs) and to total central He depletion ($Y_c$ < 1 × 10$^{-4}$) at t = 5.85 Myrs. Semiconvection mixing influences the formation and extension of the blue loops (El Eid 1995). Such a feature is absent in the more recent set of models by Limongi, Straniero & Chieffi (2000), Meynet & Maeder (2003), Chieffi, Domínguez, Hoeflich, Limongi & Straniero (2003), but similar behaviour has been obtained by Hirschi, Meynet & Maeder (2004) for a 20M$_\odot$ star as a consequence with a higher mass loss rates during He-burning. Core C ignition occurs at 5.87 Myr followed by successive Ne and O burning. The beginning of each burning phase is pointed in the HR diagram in the Figure 1. From Ne-core burning onwards, the location of the modeled star in the HR diagram does not vary because of the short timescales of the late burning stages. The last stage corresponds to a star with an envelope of 0.29 M$_\odot$ and surface abundances $X_s$ = 0.44, $Y_s$ = 0.54, and $N_s$ = 0.009. The pre-SN star has a final total mass of 9.4 M$_\odot$, with a T$_{eff}$ ~ 11,688 K and R$_{eff}$ ~ 110 R$_\odot$, resembling a BSG before SN explosion. We called this type of stars "luminous blue super giant" (LBSG) to distinguish them from normal blue super giant (BSG) of much lower masses. Using the envelope mass criterion from Heger et al. (2003), the explosion would be an SNIIb type.

**Table 1.** Evolutionary phases.

| Model | Evolutionary path |
|---|---|
| 23 M$_\odot$ | MS → RSG → SNIIp |
| 28 M$_\odot$ | MS → RSG → LBSG → SNIIb |
| 29 M$_\odot$ | MS → RSG → LBSG → YSG → SNIIb |
| 30 M$_\odot$ | MS → RSG → LBSG → WNL → WNE → WC → SNIb |
| 33 M$_\odot$ | MS → RSG → LBSG → WNL → WNE → WC/WO → SNIc? |

**Table 2.** Lifetimes of nuclear burning.

| Model | H-burning (×10$^6$ yrs) | He-burning (×10$^5$ yrs) | C-burning (yrs) | Ne-burning (yrs) |
|---|---|---|---|---|
| 23 M$_\odot$ | 6.374 | 6.01 | 7 243 | 3.64 |
| 28 M$_\odot$ | 5.252 | 5.89 | 2 238 | 1.85 |
| 29 M$_\odot$ | 5.100 | 5.45 | 2 079 | 1.19 |
| 30 M$_\odot$ | 4.959 | 5.15 | 3 286 | 0.55 |
| 33 M$_\odot$ | 4.607 | 5.10 | 3 576 | 1.29 |

### 3.3. 29 M$_\odot$ model

The evolution is quite similar to the 28 M$_\odot$ stellar model, because it loses 1.53 M$_\odot$ in the MS during 5.1 Myr and 16.37 M$_\odot$ in the RSG during 3.09 ×10$^5$ yr. After the RSG, the 29 M$_\odot$ model also performs a blue loop becoming an LBSG star. Nevertheless, the blue loop is shorter due to a higher mass loss rate in the RSG (Renzini et al. 1992). During the He shell burning the star comes back to the red side of the HR diagram igniting successively C, Ne and O in the core. The last computed model is an O-core burning star with a final total mass of ≈ 10M$_\odot$. The envelope has 0.34 M$_\odot$ composed mainly of H, He, and N ($Y_s$ = 0.52, $X_s$ = 0.46, $N_s$ = 0.008). The pre-SN star has a T$_{eff}$ ~ 7,509 K and a R$_{eff}$ ~ 277 R$_\odot$, looking like a yellow super giant (YSG) before the SN explosion.

### 3.4. 30 M$_\odot$ model

The 30 M$_\odot$ loses 1.67 M$_\odot$ in the MS during 4.959 Myr and 18.3 M$_\odot$ in the RSG phase during 2.94 ×10$^5$ yr. After RSG, the star evolves towards hotter temperatures because of the high mass loss rate. At 5.28 Myr, the star becomes a 9.9 M$_\odot$ WR star while it is burning He in the core. The WR phase starts as WR-H rich (WNL), and due to the strong mass loss rate, the star quickly loses the stellar envelope, diminishing the stellar mass and luminosity, as is notable in the HR diagram (Figure 1). During WNL, the star loses 1.15 $M_\odot$. We note that recent WR mass-loss rate estimates are lowered by a factor of 2-3 due to clumpiness (Hamman and Koesterke 1998), so our WR mass loss rates are an upper limit. During WR, the star develops three distinct phases: 1) the star spends $4.2 \times 10^4$ yr as WNL with a declining presence of surface-H. The He and N abundances begin to increase steeply due to the high mass loss rate that peels off the star and exposes layers enriched with elements from the CNO cycle. At 5.33 Myr, the He and N reaches the maximum value on surface (Figure 3). 2) The surface-H vanishes ($Y_s = 0.98$) and the star begins the WN H-poor (WNE) stage during $5.8 \times 10^4$ yr with a slightly lower mass loss rate. The WNL wind is highly enriched with He and N. The high mass loss rate causes the He and N-surface abundances to diminish when the star exposes its deeper layers processed by $3\alpha$ reactions. 3) When the surface-C abundance is larger than 3 % in mass, a WC phase begins. During the WC phase, the stellar surface temperature remains almost constant and the luminosity decreases. Core He-exhaustion corresponds to the luminosity minimum. At the ignition of an He-burning shell, the stellar luminosity increases slightly and the star returns to the red side of the HR diagram with lower luminosity. During the redward evolution in the HR diagram, the star ignites C, Ne, and O-core burning. The pre-SN model is a 5.9 $M_\odot$ star with a 1.6 $M_\odot$ Si/O core. The star loses the H envelope and it shows the He core partially processed by 3 $\alpha$ reactions ($Y_s = 0.73$ and $C_{sup} = 0.18$). The stellar radius is $R_{eff} \approx 3$ $R_\odot$ and the stellar surface has a $T_{eff} \approx 56,800$ K. Here, the stellar radius or "hydrostatic radius" of a WR star is defined as the radius of the sonic point in the outflow (Heger & Langer 1996). The presence of surface-He indicates that the SN light curve would be an SNIb type.

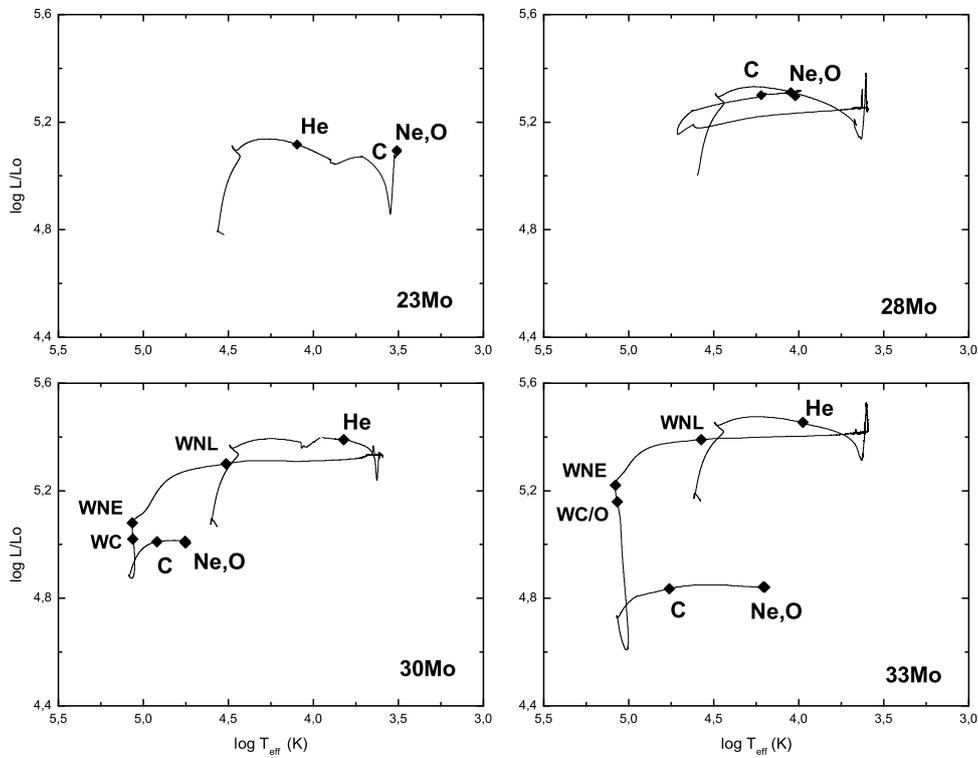

**Fig. 1.** Evolutionary tracks for the models with 23 $M_\odot$ (top left), 28 $M_\odot$ (top bottom), 30 $M_\odot$ (bottom left), and 33 $M_\odot$ (bottom right) from the ZAMS to the pre-SN stage. The rhombuses mark the beginning of each core-burning phase. For the 28 $M_\odot$ model, the He-core burning is not included since it lies at the same position of the Ne-burning phase.

### 3.5. 33 $M_\odot$ model

This stellar model loses 2.16 $M_\odot$ in the main sequence during 4.607 Myr. As an RSG, the star spends $1.7 \times 10^5$ yr losing 18.9 $M_\odot$. In a similar way to the 30 $M_\odot$ star, this model also becomes a WR and the transition occurs at $\approx 4.8$ Myr. Similar to the previous case, the whole WR phase is divided into three stages. The WNL (H-rich) stage lasts $3.6 \times 10^4$ yr and the star loses almost 1 $M_\odot$ in a WR wind. During WNL, the surface temperature of the star increases steeply reaching values of 120,000 K when the star peels off its hydrogen envelope ($Y_s = 0.98$) starting the WNE (H-poor) stage. The mass-loss rate is reduced slightly and the star loses 0.8 $M_\odot$ in an He and N-rich wind, as is shown in Figure 3. The He and N abundances reach their maximum values before the WNE (H-poor) stage. The strong mass loss exposes the $3\alpha$ processed material and star becomes a WC/WO star when the star peels off deeper layers. The strong mass-loss rate peels off the star so deep that we observe a larger amount of material processed by $3\alpha$ reactions than the one observed in the 30 $M_\odot$ model. In the WC/WO stage, the star evolves practically at a constant surface temperature and the luminosity decreases. Shortly before C core ignition, a He-burning shell expands the stellar envelope and the star comes back to the red side of the HR diagram, igniting C, Ne, and O in a successive way. Our last model has an Si/O core of 1.6 $M_\odot$, and it is close to SN explosion, only a few days or hours to ignite Si. The intense WR mass loss has completely removed the stellar envelope and the star is exposing the He burning shell where most of He has been processed by $3\alpha$ reaction. The C-surface abundance is higher than He or O. There is no presence of either H or N in the stellar surface (the N has been destroyed at the high temperatures of the $3\alpha$ reaction). The pre-SN star is a WC with 4.57 $M_\odot$ and $T_{eff}$ = 15,000 K. The absence of H and the low He-surface abundance indicate that the most probable SN type would be an SNIc.

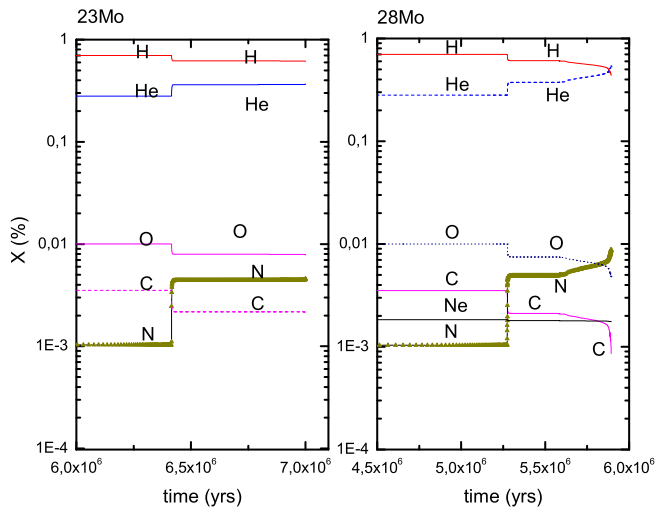
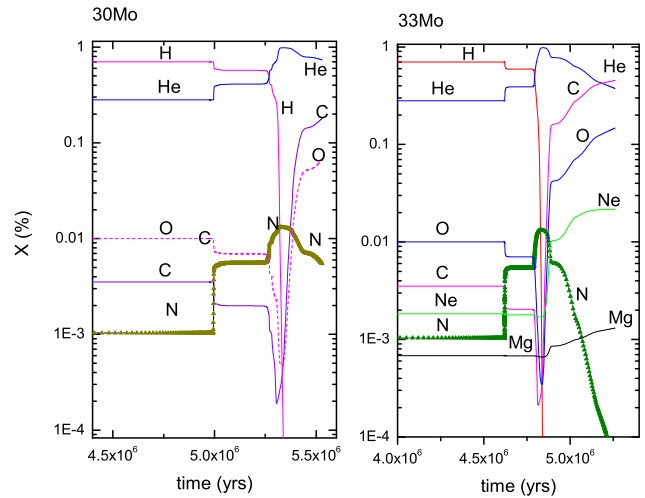

**Fig. 2.** Surface mass fractions of elements as function of time in the 23 $M_\odot$ and 28 $M_\odot$ models. During MS the surface abundances have the same value. Abundances of all isotopes of same element are grouped in a tag to compare them with Cas A composition; i.e. all carbon isotopes are grouped in the "C" abundance tag.

**Fig. 3.** Surface mass fractions of elements as function of time in the 30 $M_\odot$ and 33 $M_\odot$ models. During MS, the surface abundances have the same value. Abundances of all isotopes of same element are grouped in a tag to compare them with Cas A composition; i.e. all carbon isotopes are grouped in the "C" abundance tag.

**Table 3.** Final stellar models.
**Table 4.** $M_f$ refers to final stellar mass, $M_{He}$ is He-core mass, $M_{CO}$ is CO-core mass and $M_{env}$ is the stellar envelope mass, all in units of $M_\odot$. $R_{fin}$ is the final stellar radius.

| Model | $M_f$ ($M_\odot$) | $M_{He}$ ($M_\odot$) | $M_{CO}$ ($M_\odot$) | $M_{env}$ ($M_\odot$) | $R_{fin}$ ($R_\odot$) | $T_{eff}$ (K) |
|---|---|---|---|---|---|---|
| 23 $M_\odot$ | 13.2 | 6.45 | 2.23 | 6.75 | 700 | 4 111 |
| 28 $M_\odot$ | 9.3 | 8.97 | 3.37 | 0.29 | 100 | 11 688 |
| 29 $M_\odot$ | 9.9 | 9.31 | 3.07 | 0.34 | 270 | 7 509 |
| 30 $M_\odot$ | 5.8 | 5.85 | 2.86 | – | 3 | 57 430 |
| 33 $M_\odot$ | 4.57 | 4.57 | 2.93 | – | 36 | 15 719 |

Table 1 shows a summary of the evolutionary sequences of our stellar models, including the SN type associated to each star. The SN type was determined using the stellar envelope mass criterion according to Heger et al. (2003). There is no model ending as WNL/WNE due to the high mass-loss rates at WR stages. The WR mass loss rates used here are slightly higher than those in Nugis & Lamers (2000), and the effect of clumpiness have not been taken into account, thus, the WR mass loss rates can be considered as upper limits.

Table 2 shows the nuclear burning lifetimes for each star. Core He-exhaustion corresponds to the minimum in luminosity (similar to Figure 2 by Woosley et al. 1995). The subsequent expansion is driven by shell helium burning. Although burning lifetimes after C-ignition are short, the final evolution in the HR diagram for 30 $M_\odot$ and 33$M_\odot$ produce redward excursions. The redward excursion of this type is an anomalous feature, as our referee pointed out, and none of the models avaliable in the literature show such behavior. We think that the reason for *"life after C-ignition"* in the HR diagram of these models is the rather low final mass (6 $M_\odot$ or less), giving long Kelvin-Helmholtz times (7 $\times 10^3$ yrs in 30 $M_\odot$ and 5 $\times 10^3$ yrs in 33$M_\odot$ models at C-burning ignition, even larger than C-burning lifetimes). This anomalous behavior will be the subject of a future study to confirm our finding. Table 3 shows a summary of the main characteristics of our pre-SN models.

## 4. Hydrodynamical simulations

### 4.1. Main sequence and red super giant phases

The CSM evolution strongly depends on stellar evolution out through the stellar wind properties (wind momentum and mechanical energy). In this study, we do not consider the effects of photoionization from stellar radiation, magnetic fields, or heat conduction. To simulate the ionization post-RSG from the stellar radiation field, we change the lower cutoff temperature to $10^4$ K, saving a large amount of computational time (GLM96AB). By doing this, we assume that the wind and RSG shell will be ionized (optically thin), and its temperature corresponds to the ionization equilibrium temperature.

For simplicity, the numerical calculation are divided into four stages: MS (1D), RSG (1D), blue loop or WR star (2D), and SN interaction (2D). Since our main focus is the final SN

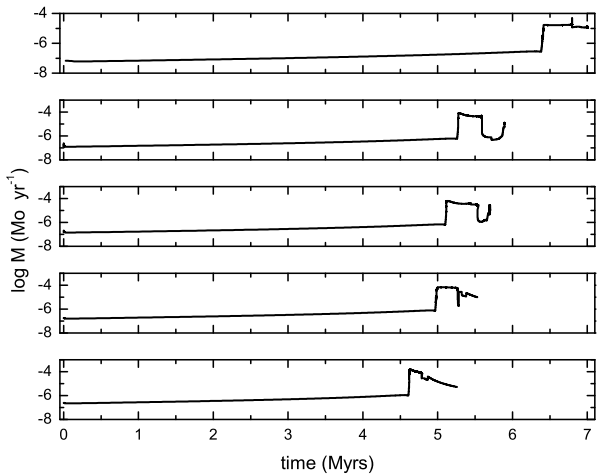

**Fig. 4.** Logarithm of the mass loss rate as function of time (from top to bottom) for the 23, 28, 29, 30 and 33 M$_\odot$ models. The MS, RSG and post-RSG stages can be clearly distinguished. The highest mass loss rates corresponds to RSG stages.

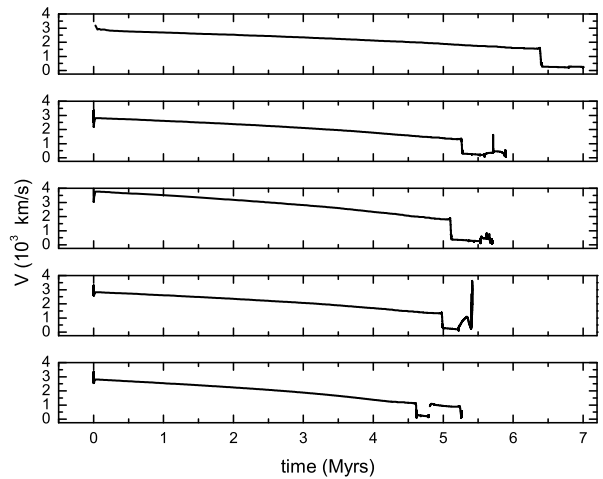

**Fig. 5.** Terminal wind velocities as a function of time (from top to bottom) for the 23, 28, 29, 30 and 33 M$_\odot$ models. The lowest wind velocities corresponds to RSG stages.

interaction with the nearby CSM, the HII region around the MS is not computed either.

Figures 4 and 5 show the stellar mass-loss rates and wind velocities as a function of time obtained from the stellar evolution calculations. The average wind velocities, total mass lost, and lifetimes at each stages are compiled in Tables 3, 4, 5, 6, and 7.

Figures 4 and 5 reflect the different stellar evolutionary stages. During MS a slightly increase in the mass-loss rate is observed as a function of effective stellar radius and temperature. The opposite behavior for the velocity. During RSG, the mass-loss rate increases nearly by two orders of magnitude. The RSG wind is much slower and denser than the MS wind.

The hydrodynamical calculations start at ZAMS, in a homogeneous and quiescent ambient medium with an uniform density appropriate for Cas A of $n_o = 13 cm^{-3}$ (Willingale et al. 2003), and an initial thermal energy density of $1 \times 10^{-13} erg/cm^3$.

During the MS, the wind kinetic energy is converted into thermal energy by collision at the reverse shock, forming an adiabatic bubble surrounded by a shocked, swept-up shell of interstellar material (MS shell) (Weaver et al. 1977). Using an average value of $L_w = 5.5 \times 10^{35} ergs^{-1}$ for the mechanical luminosity and $n_o = 13 cm^{-3}$ for the ISM density in the Weaver et. (1977) analytical solutions, we obtain MS shell radii in close agreement with the calculations (Fig. 6).

During the RSG phase, there is a large uncertainty in the wind velocity (Chiosi & Maeder 1986). In our case, the stellar wind-velocity outputs from the stellar evolution code decreases down to 250 km s$^{-1}$, although we would expect RSG wind velocities to be even lower. Because of the uncertainty on RSG wind velocities, ranging between 10 - 160 Kms$^{-1}$, and the lack of theoretical values (Dupree 1986), we used a standard velocity of 10 Kms$^{-1}$, in agreement with other authors (Chevalier and Oishi 2003).

We performed the CSM evolution for RSG phase in 1D, since the RSG shell is quite stable for low RSG wind velocities (García-Segura et al. 1996A, Freyer et al. 2006). Faster winds are subject to the Rayleigh-Taylor (R-T) instability (for LBVs see García-Segura et al. 1996B; for RSG see Dwarkadas 2007).

The R-T instability produces fingers in the RSG shells, but the instability does not affect the position of their inner edge.

The duration of the RSG stage is different for each star, ranging from $1 \times 10^5$ to $6 \times 10^5$ yr, but the impact in the CSM is quite similar. During RSG, the winds build up shocked shells where the ram pressure of the RSG wind balances the thermal pressure of the main sequence bubbles. The CSM density profiles for each star at the end of the RSG phases is shown in Fig. 6. We observe that larger RSG shells correspond to low masses since they spend longer times as RSGs. Denser winds correspond to more massive stars. On the other hand, MS shell radii are quite similar at the end of RSG phases, since higher stellar masses with higher mechanical luminosities live shorter.

### 4.2. Post-RSG evolution and SN blast wave

#### 4.2.1. 23 M$_\odot$ model. RSG progenitor

The 23 M$_\odot$ model ends up as a 13.2M$_\odot$ RSG SN progenitor. At the time of SN explosion, the CSM is composed of a free-streaming and dense RSG wind with a density profile $\rho \propto r^{-2}$ surrounded by an RSG-shocked wind shell. A hot MS bubble remains outside the RSG shell, delimited by an ISM shocked shell at $\sim 33$ pc (MS shell). In this medium, we computed the ejection of mass from SN explosion in a two-dimensional numerical grid. The simulation was done in a numerical grid of $500 \times 300$, covering the surrounding space from the star up to 15 pc, in order to compare with the size of Cas A. The numerical domain covers 15 pc $\times$ 30° with a resolution of 0.03 pc in radial direction and 0.1° for the azimuthal direction.

To study the blast wave interaction with the CSM, we assume an energy of $2 \times 10^{51} erg$ for the SN explosion appropriated for Cas A (Hwang & Laming 2003; Willingale et al. 2003; Vink et al. 2004). The outer stellar layers are ejected in the explosion (11.6M$_\odot$) leaving a stellar remnant of 1.6M$_\odot$. The remnant mass was chosen according to deleptonized cores in massive stars (Woosley et al. 2002). For simplicity, we set up the explosion with an over-pressure region at the center of the grid, with constant density (corresponding to the ejecta mass over the volume region) and a high thermal energy (= $10^{51}$ erg) and with an initial velocity assuming that $10^{51}$ erg were in the form of kinetic energy. The approximation is sufficient for the scope of this study. The density profile, after several years, is determined by the hydrodynamical expansion of the ejecta. Using numerical inspection, we observed a density profile of $\rho \propto r^{-10.2}$ at 167 yr after explosion, very similar to those proposed by Matzner and McKee (1999) for this SN type.

**Table 5.** Wind properties of the 23 $M_\odot$ model. Quantities are obtained from STERN code and refers to average values. The (*) indicate that the reported value was changed in the hydrodynamical simulations (see text for details).

| Stage | $\Delta t$ (Myrs) | log $\dot{M}$ ($M_\odot$/yr) | $\Delta M$ ($M_\odot$) | $v_\infty$ (km s$^{-1}$) |
|---|---|---|---|---|
| MS | 6.374 | -6.87 | 0.81 | 2,000 |
| RSG | 0.622 | -4.77 | 8.96 | 260 (*) |

**Table 6.** Wind properties of the 28 $M_\odot$ model.

| Stage | $\Delta t$ (Myrs) | log $\dot{M}$ ($M_\odot$/yr) | $\Delta M$ ($M_\odot$) | $v_\infty$ (km s$^{-1}$) |
|---|---|---|---|---|
| MS | 5.252 | -6.55 | 1.39 | 2,000 |
| RSG | 0.313 | -4.27 | 16.85 | 260 (*) |
| LBSG | 0.303 | -5.96 | 0.33 | 400 |

**Table 7.** Wind properties of the 29 $M_\odot$ model.

| Stage | $\Delta t$ (Myrs) | log $\dot{M}$ ($M_\odot$/yr) | $\Delta M$ ($M_\odot$) | $v_\infty$ (km s$^{-1}$) |
|---|---|---|---|---|
| MS | 5.100 | -6.48 | 1.53 | 2,000 |
| RSG | 0.309 | -4.28 | 16.37 | 260 (*) |
| LBSG | 0.218 | -6.28 | 0.27 | 600 |

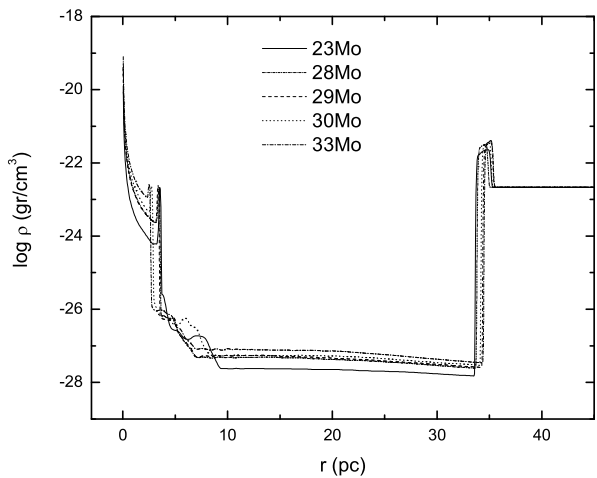

**Fig. 6.** Logarithm of gas density at the end of each RSG phases. The swept up shells formed at the main sequence phases are located at distances of ∼ 35 pc. The RSG pilled up shells (∼ 3 pc) form inside of the MS bubbles, where the RSG wind ram pressure balance the thermal pressure of the main sequence bubble. The largest RSG shell radius correspond to the 23 $M_\odot$ model.

Figure 7 shows the logarithm of density snapshots for eight different times, from the explosion up to 1167 yr after it. The first slice on the left corresponds to the pre-SN CSM ($t_{SN}$ = 0 yr), and the hydrodynamical evolution runs clockwise. The SN ejecta interacts first with 5.8 $M_\odot$ of dense RSG wind ($\rho \propto r^{-2}$). After that, the ejecta encounters a 3.14 $M_\odot$ RSG shell at a distance of 3.5 pc from the star, with a low expansion velocity (5 km s$^{-1}$) with respect to the star. The initial velocity of the forward shock is 4,200 km s$^{-1}$ and the ejected mass is 11.6 $M_\odot$. The slice 2 shows the density at 334 yr after the explosion (the age of Cas A). At this moment, the forward shock has a v ∼ 4,000 km s$^{-1}$, and it has not interacted yet with the RSG shell. The forward shock front is at 2.2 pc and it has shocked 3.6 $M_\odot$ of the RSG stellar wind. The blast wave hit the RSG shell 830 yr after the explosion.

#### 4.2.2. 28 and 29 $M_\odot$ models. LBSG progenitors

The hydrodynamical evolution of both modeled stars is quite similar, so we only discuss the 28 $M_\odot$ model. After the RSG phase, the star evolves blueward performing a blue loop. During the blue loop, the wind velocity increases to reach a maximum value of 550 km s$^{-1}$ according to empirical LBSG wind velocities (Chevalier and Fransson 1987). The interaction of the supersonic LBSG wind with the slow and dense RSG wind material produces a double shock-front structure, and a new bubble of hot shocked wind (*LBSG bubble*). The thermal pressure of the LBSG bubble pushes the medium in a quasi-adiabatic shock and sets up a new shell of shocked, swept-up RSG wind (*LBSG shell*). Figure 8 shows the formation of the LBSG shell through the blue loop. The LBSG shell is affected by Rayleigh-Taylor (RT) instabilities at the interface, when it is accelerated by the LBSG bubble pressure. The LBSG shell collides with the RSG shell at 3.4 pc. The expansion velocity of the LBSG shell (35 km s$^{-1}$) is higher than the one of RSG shell (5 km s$^{-1}$), and the LBSG and RSG shell remain separated for ∼ $10^5$ yr previous to their collision. The peak density of the LBSG shell is 2.6 times more than the RSG shell. Both shells merge and form a massive shell moving out with a common velocity of ≈ 18 km s$^{-1}$. The new LBSG+RSG shell moves outwards into the MS bubble, and it develops RT instabilities in the outer edge, since the shell is decelerating. The outer edge is subsonic with respect to the MS bubble, and there is not a forward shock. As the shell propagates outwards eventually breaks-out (Fig. 8, slice 5).

**Table 8.** Wind properties of the 30 $M_\odot$ model.

| Stage | $\Delta t$ (Mysr) | log $\dot{M}$ ($M_\odot$/yr) | $\Delta M$ ($M_\odot$) | $v_\infty$ ( km s$^{-1}$ ) |
|---|---|---|---|---|
| MS | 4.959 | -6.34 | 1.67 | 2,500 |
| RSG | 0.294 | -4.19 | 18.30 | 250 (*) |
| WNL | 0.042 | -4.55 | 1.15 | 1,000 |
| WNE | 0.058 | -4.89 | 0.75 | 1,000 |
| WC/WO | 0.149 | -4.80 | 2.23 | 1,000 (*) |

**Table 9.** Wind properties of the 33 $M_\odot$ model.

| Stage | $\Delta t$ (Myrs) | log $\dot{M}$ ($M_\odot$/yr) | $\Delta M$ ($M_\odot$) | $v_\infty$ ( km s$^{-1}$ ) |
|---|---|---|---|---|
| MS | 4.607 | -6.39 | 2.16 | 2,500 |
| RSG | 0.172 | -3.96 | 18.90 | 250 (*) |
| WNL | 0.036 | -4.55 | 1.00 | 1,000 |
| WNE | 0.036 | -4.66 | 5.46 | 1,000 |
| WC/WO | 0.394 | -5.00 | 0.40 | 1,000 (*) |

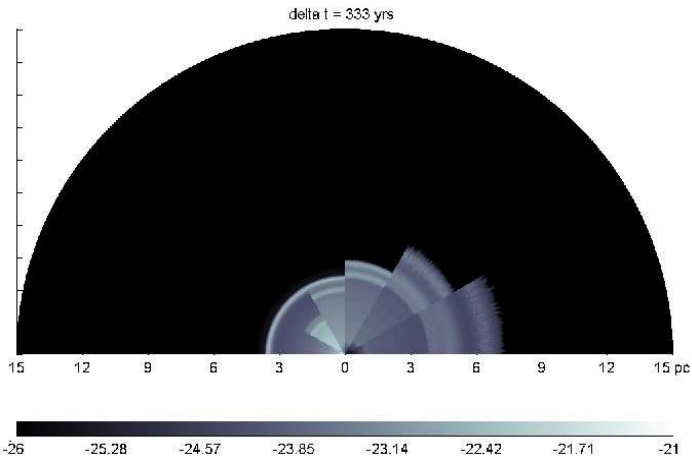

**Fig. 7.** Blast wave evolution for the 23 $M_\odot$ model (logarithm of gas density in g cm$^{-3}$). The pre-SN CSM is shown on the left slice. Time runs clockwise, with 333 yr time difference between slices. Slice 2 corresponds to the age of Cas A. The supernova blast wave as well as the result from the collision are R-T unstable, but this is not very well resolved at this resolution.

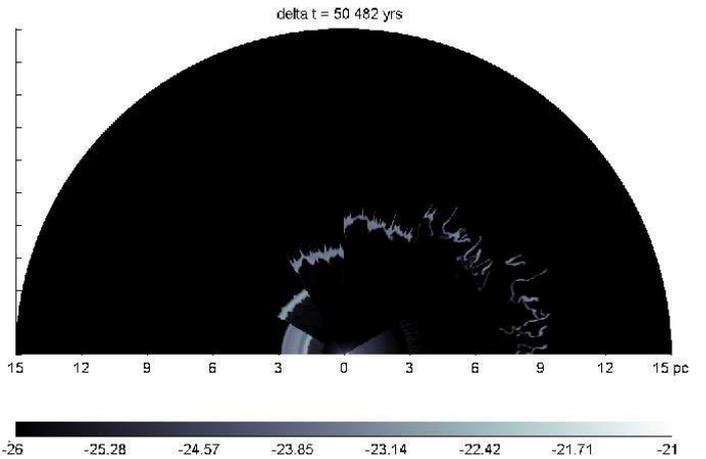

**Fig. 8.** Post-RSG CSM evolution of the 28 $M_\odot$ model. The figure shows the logarithm of gas density (g cm$^{-3}$). The CSM at the onset of LBSG phase (t = $5.64 \times 10^6$ yr) is shown on the left slice. The evolution runs clockwise. Slices have a $\Delta t$ = 50,482 yr. Slice 6 occurs at t = $5.893 \times 10^6$ yr, when the star is at core O exhaustion (pre-SN).

The final position of the merged shell is spread between 7.6 - 9 pc. It has a swept-up mass of 17.2 $M_\odot$ and some regions have a peak density higher than $\rho \approx 1.67 \times 10^{-24}$ gr/cm$^3$, able to be detected in circumstellar nebula (GLM96A.) The mass of the wind inside the LBSG bubble is less than 0.05 $M_\odot$. The density profile of the pre-SN medium around the star of 28 $M_\odot$ is shown in slice 8 of Figure 8. The LBSG shell is composed mainly of H, He, and N because it was formed from swept-up RSG wind. The LBSG wind is enhanced with He and N, and it is H deficient, although H still is an important wind component. The stellar surface of the pre-SN LBSG is He, H and N abundant in that order.

The blast wave was simulated like in the 23 $M_\odot$ model, using the same SN energy. In this case, the SN ejects a mass of 7.86 $M_\odot$, leaving a stellar remnant of 1.5 $M_\odot$. The blast wave evolution is shown in Figure 9. The initial forward shock velocity is 8,000 km s$^{-1}$. In slice 2 of Figure 9 we observe the CSM at the age of Cas A ($t_{SN}$ = 334 yr). Here, the blast wave has shocked all the free LBSG wind and starts the interaction with the LBSG shocked wind (LBSG bubble). The forward shock is at 3.3 pc and has shocked less than 0.1 M$_\odot$ of the LBSG stellar wind. The collision of the blast wave with the LBSG bubble produces a direct shock front propagating into the bubble and a reflected shock traveling back. The collision causes a sudden increase in the temperature of the shocked gas and an increment in the x-ray emission.

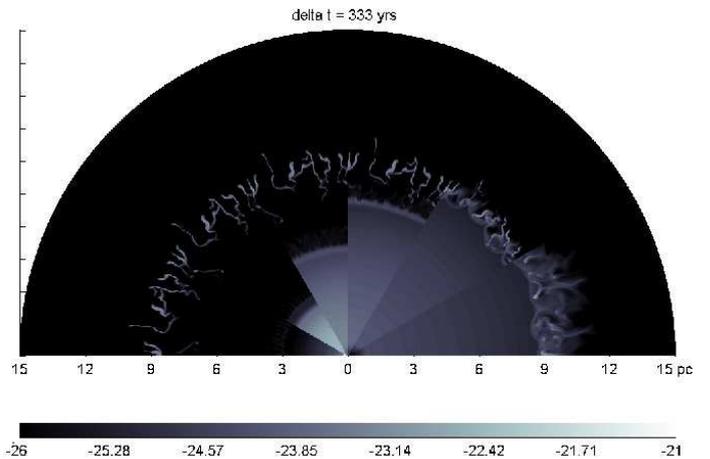

**Fig. 9.** Blast wave evolution for the 28 $M_\odot$ model (logarithm of gas density in g cm$^{-3}$). The pre-SN CSM is shown on the left slice. Time runs clockwise, with 333 yr time difference between slices. Slice 2 corresponds to the age of Cas A.

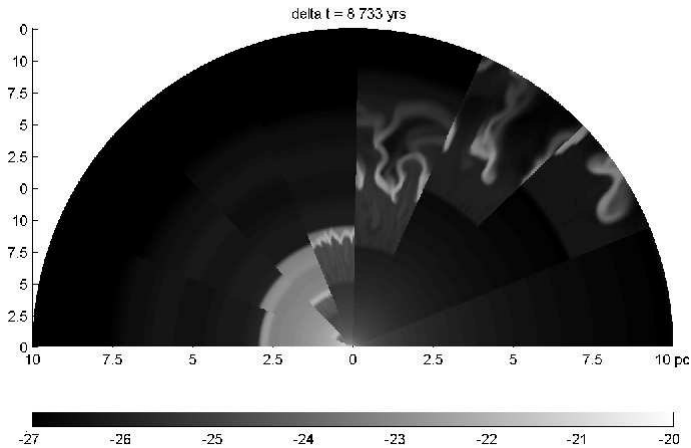

**Fig. 10.** Post-RSG CSM evolution of the 30 M$_\odot$ model. The figure shows the logarithm of gas density (g cm$^{-3}$). The CSM at the end of RSG phase (t = 5.267 × 10$^6$ yr) is shown on the left slice. The evolution runs clockwise. Slices have a $\Delta t$ = 8,733 yr. The SN occurs 1.3 × 10$^5$ yr after slice 8.

The direct (forward) shock collides with the merged (LBSG+RSG) shell at 1,100 yr after the explosion, with a velocity of 6,500 km s$^{-1}$. The collision transmits a shock into the merged (LBSG+RSG) shell, with a strong acceleration of the shell. The shock loses a large amount of energy at the collision. At the collision knots traveling outwards are formed with a wide range of velocities and densities.

### 4.2.3. 30 and 33 M$_\odot$ models. WR progenitors

After the RSG phases, the 30 and 33 M$_\odot$ models shed their H envelope and become WR stars with faster winds (> 1,000 km s$^{-1}$) and with lower densities than the RSG winds. The chemical composition in the WR winds is mainly He, H and N, in the WNL phase with a declining presence of H. WNE winds are composed mainly of He and N. The WC/O winds are still He-rich but with a strong abundance of C and O and declining presence of N (Figure 3).

At the beginning of the WR phases, the RSG shells are located at 2.5 pc. They are ionized by the intense stellar radiation coming from the WR stars and their temperature increases to the photoionization equilibrium temperature (T ∼ 10$^4$ K). Fast WR winds sweep up the dense and slow RSG winds to build up swept-up shells (*WR shells*). Figure 10 shows the logarithm of gas density at the end of the RSG phase (slice 1) and the formation of the WR shell for the case of the 30 M$_\odot$ model. The expansion velocity of the WR shell is ∼ 100 km s$^{-1}$. The WR shell hits the RSG shell at 5.3 Myr, 2.2 ×10$^5$ yr before SN explosion. As a result of the strong collision, both shells are broken at radii smaller than 5 pc (slice 5 of Fig. 10), forming clumps and tails of dragged shell material. At SN explosion, the fragmented shell consists of dense knots lying at a distance greater than 20 pc and irregular tails that are spread in an annulus of 20 pc of inner boundary and 30-35 pc in the outer edge (the MS shell at ∼ 35 pc).

The blast waves were simulated similar to the 23 and 28 M$_\odot$ models. The modeled WR stars explode with an energy of 2 × 10$^{51}$ erg, ejecting into space less than 5 M$_\odot$, leaving stellar remnants of ∼ 1.5 M$_\odot$ (Table 10).

Initial velocities of the SN forward shocks are ∼ 7,000 km s$^{-1}$, and at the age of Cas A, they have slightly decayed to 6,800 km s$^{-1}$, at distances of 3.1 pc from the explosion centers. At this age, the SN shock fronts are still interacting with the free WR winds.

The wind termination shocks of the WR bubbles are at distances greater than 10 pc, and the SN shocks take more than 600 yr to reach them. The blast waves hit the knots from the broken WR shells at times longer than 1,000 yr after the explosions.

Since the purpose of this work is to compare our simulations with the observational constraints of Cas A, we did not carry out the whole interaction of SN shock fronts with WR CSM. Interactions of SN shock fronts with WR bubbles has been simulated by Dwarkadas (2007).

## 5. Comparison with Cas A

There is a wide variety of observations to constraint the progenitor of Cas A. In this section we have used the strongest observational constraints to compare them with our stellar and hydrodynamical simulations in order to find the best progenitor model for Cas A.

### 5.1. Comparison with stellar models. Quasi Stationary Floculii

Chevalier and Kirschner (1978) report N and He excess in QSFs with ratios of N/H ∼ 7-10 and He/H ∼ 3 -5 times the solar values. The overabundances lead to proposing that QSFs were stellar material processed by the CNO cycle, transported to the surface and lost by the stellar wind (Kamper and van den Bergh 1978), or as clumpy ejecta (Chevalier & Oishi 2003). Some authors have proposed that Cas A progenitor was a WN star (Peimbert 1971, Fesen et al. 1987, Willingale et al. 2003) from the QSFs chemical ratios.

Figure **??** show the evolution of the index 12 + log(N/H) and 12 + log(He/H) on the stellar surface of the 30 and 33M$_\odot$ modeled stars to compare them with the observational ratios found in QSFs. The stellar surfaces exhibit N/H and He/H ratios similar to those in QSFs at t = 5.321 Myr for the 30 M$_\odot$ model, and at t = 4.789 Myr for the 33 M$_\odot$. In both WR stars, we observe that the QSF ratios appear at the onset of the WNE phase, so it is quite reasonable that QSFs come from WNE stellar winds. After the WNE, the N/H and He/H ratios continue increasing, and any mass ejections from the stellar surfaces after WNE will contain at least QSF ratios, or even higher.

The 23, 28, and 29 M$_\odot$ models do not exhibit surface QSF ratios at any time. The N/H and He/H ratios are always much lower than those values observed in QSFs. If we accept the scenario in which QSFs were expelled from the stellar surface as wind or blobs, then it is difficult to postulate a RSG precursor for Cas A. However, this cannot be ruled out, since stronger N enrichment is expected at the surface of rotating stars (Heger & Langer 2000).

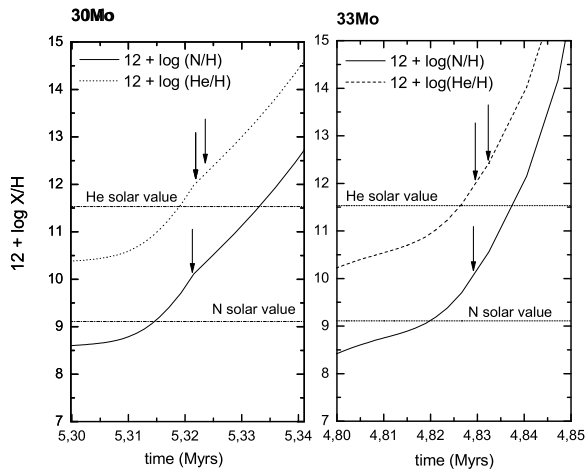

**Fig. 11.** Evolution of the 12 + log(N/H) ratio (solid line) and 12 + log(He/H) ratio (pointed line for 30M$_\odot$ model and dashed line for 33 M$_\odot$ model) on the stellar surface. We show the time interval for which the stellar surface of our models exhibits QSF ratios (from Chevalier and Kirshner 1978) marked by arrows.

### 5.2. Comparison with stellar models. Nitrogen Knots

The NKs were identified for the first time in spectroscopic observations of the outer optical knots by Fesen (2001). Fesen found four dozen fast knots ($\sim$ 8,000 - 10,000 km s$^{-1}$ ) with nitrogen-rich abundances with typical limit of N/H $\sim$ 30 times solar. Their estimated space velocities suggest a nearly isotropic ($\sim$ 10,000 km s$^{-1}$ ) ejection velocity. Because of the isotropic ejection, the N-rich ejecta fragments were presumably shrapnel from an N-rich envelope of the precursor. The presence of an appreciable H emission in just 3 of the 50 NKs suggest an H-poor photosphere when the SN occurs, since a significant mass accretion of circumstellar material onto the NKs seems unlikely. The H + N rich photospheric layers must have had very low mass (Fesen 2001), suggesting a pre-SN star whose mass loss was not enough to expose the CO layers processed by 3$\alpha$ reaction, where N is destroyed.

The 23 M$_\odot$ model has a pre-SN star that is an RSG with a very massive H envelope ($X_s$ = 0.62, $Y_s$ = 0.36). The surface N is the fourth element in abundance by mass. This H envelope is unlikely to produce and eject H poor clumps like the NKs.

The 28 and 29 M$_\odot$ models are LBSGs when they explode, and have pre-SN envelopes of small masses ($\sim$ 0.3M$_\odot$ ) composed mainly of He, H, and N (in that order by mass abundances). The nitrogen on the surface is an efficient coolant that quickly cools down the gas and forms knots resembling NKs during SN ejection. The N/H ratio found in NKs comes from material that has undergone CNO processing up to $>$ 90 % of the H-depleted (Young et al. 2005), and the material in our LBSG pre-SN envelopes are $\sim$ 60 % H-depleted. Then, our LBSG progenitors are the most likely stars to produce clumps resembling NKs.

The pre-SN of the 30 M$_\odot$ model has N on its surface, but does not have a significant stellar envelope. During ejection, the N on the surface could cool down to form knots at high velocities, but lacking hydrogen. In the 33 M$_\odot$ model, the WR mass loss removes all of the N-rich material from the stellar surface. The peeling off reaches deep layers where most of the material was processed by the 3$\alpha$ nuclear reaction, destroying the N, so the ejected mass is unable to form NKs.

### 5.3. Comparison with stellar models. SN ejected mass

The mass ejected by the SN explosion of Cas A has been measured using several methods. Willingale et al. (2002) combine x-ray spectral line fitting with emission models to estimate the total amount of emitting material, estimating an ejected mass of 2 - 4 M$_\odot$ with 1-3 M$_\odot$ of oxygen. Vink et al. (1996) has determined an ejected mass of $\sim$ 4M$_\odot$ with an ejected oxygen mass of $\sim$ 2.6M$_\odot$ .

The final masses of our stellar models are shown in Table 3, and the total ejected masses and yields by elements are shown in Table 10, using mass cuts according to Woosley et al. (2002). Ejected masses by the 23, 28, and 29 M$_\odot$ models are higher than the observational values in Cas A. The RSG progenitor (23 M$_\odot$ model) ejects the highest amount of mass, 11.6 M$_\odot$, and the LBSG progenitors (28 and 29 M$_\odot$ models) ejects $\sim$ 8 M$_\odot$. Neither the RSG nor the LBSG models satisfy the low ejecta mass constraint.

To eject the mass observed in Cas A (2 - 4 M$_\odot$ ), our RSG and LBSG stars must lose a large amount of material previous to the SN explosion. Higher mass-loss rates in RSG have been proposed recently (Schroder & Cuntz 2005) that could reduce the final mass of our RSG and LBSG precursors. However, a stronger mass loss in RSG phases will modify the stellar evolution, actually forming WR stars. The ejected mass is also reduced if the stellar explosion leaves a black hole as a remnant, with $\geq$ 6 M$_\odot$ . The stellar remnant of Cas A was observed in x-rays (Tananbaum 1999), whose nature is still being discussed (Mereghetti et al. 2002). However, stars with inital masses lower than 25 M$_\odot$ seem unlikely to produce black holes (Heger et al. 2003).

Mass-loss rates in WR stars are high enough to uncover the CO-rich material, reducing final masses to $<$ 6 M$_\odot$. An ejected mass of 2-4 M$_\odot$ leaves a stellar remnant of 1-2 M$_\odot$ and would produce a neutron-star remnant. Ejected masses by our WR precursors lie in the range of 3-4.5 M$_\odot$ , in excellent agreement with the values measured in Cas A.

### 5.4. Comparison with hydrodynamical models. Forward and reverse shock fronts

The forward shock of Cas A has been detected in x-rays at a radius of $\sim$ 150″ $\pm$ 12″ (2.5 $\pm$ 0.2 pc), interacting with the CSM (Gotthelf et al. 2001). This shock front has a velocity of $\sim$ 6,000 - 7,000 km s$^{-1}$ (DeLaney and Rudnick 2003). The reverse shock position is less clear, at an average radius of 1.6 pc with a variation of 14 % around the remnant (Gotthelf et al. 2001).

The CSM around our RSG progenitor is a dense and uniform, free expanding wind with a density profile of the type $\rho \propto r^{-2}$, extended up to 3.5pc, surrounded by a dense RSG, piled up shell. Beyond the RSG shell is the hot bubble and MS shell at 33pc. Initially, the computed blast wave expands at $\sim$ 4,200 km s$^{-1}$ . At 334 yr, the SN forward shock is located at 2.2 pc, still inside of the RSG wind, with an expansion velocity of $\sim$ 4,000 km s$^{-1}$ . The SN reverse shock is at 1.7 pc , giving a ratio of 1.3 for their radii. The observed ratio for the radii of Cas A is $\sim$ 1.5. The self-similar solutions of Chevalier (1982) show that an ejecta of the type $\rho \propto r^{-10.12}$, interacting with a stellar wind, produce a ratio of 1.26. Thus, the radii of both shocks and their ratio are in close agreement with those observed in Cas A, in favor of the 23 M$_\odot$ model.

**Table 10.** Ejected masses from the stellar models. M$_{rem}$ is the mass cut chosen from Woosley et al. (2002). M$_{ej}$ is total ejected mass. Last five columns show the yields by element.

| Model | M$_{rem}$ (M$_\odot$) | M$_{ej}$ (M$_\odot$) | H (M$_\odot$) | He (M$_\odot$) | C (M$_\odot$) | N (M$_\odot$) | O (M$_\odot$) |
|---|---|---|---|---|---|---|---|
| 23 M$_\odot$ | 1.60 | 11.6 | 3.84 | 6.08 | 0.57 | 4.1 ×10$^{-2}$ | 0.93 |
| 28 M$_\odot$ | 1.50 | 7.86 | 0.21 | 3.14 | 1.38 | 1.4 ×10$^{-2}$ | 2.77 |
| 29 M$_\odot$ | 1.53 | 8.42 | 0.10 | 3.39 | 1.50 | 2.0 ×10$^{-2}$ | 3.13 |
| 30 M$_\odot$ | 1.45 | 4.40 | 2.3 ×10$^{-9}$ | 1.51 | 0.99 | 3.9 ×10$^{-3}$ | 1.51 |
| 33 M$_\odot$ | 1.50 | 3.12 | 1.3 ×10$^{-9}$ | 0.43 | 0.92 | 1.1 ×10$^{-5}$ | 1.26 |

The pre-SN CSM around the LBSG progenitor also has a density profile $\rho \propto r^{-2}$, but less dense than the RSG medium because of the higher wind velocity. The free LBSG wind has a mass below 0.1 M$_\odot$ and the termination shock is at ∼ 1 pc. The shocked LBSG wind builds up a hot LBSG bubble, surrounded by a dense, piled-up "LBSG+RSG" shell at ∼ 7.6 - 9 pc from the star. The "LBSG+RSG" shell has a mass of 17.2 M$_\odot$. Most of the mass in the shell has a peak density higher than $1.67 \times 10^{-24}$ g cm$^{-3}$, which can be detected in the CSM (GLM96A). For this case, the blast wave has an initial v ∼ 8,000 km s$^{-1}$ and the shock velocity does not change appreciably during the LBSG wind interaction, due to the low density. At 334 yr, the SN forward shock is at 3.3 pc, with v ∼ 7,000 kms$^{-1}$, and the reverse shock is at 2.4 pc, giving a ratio of r$_f$/r$_r$ ∼ 1.4, quite similar to the value observed in Cas A. However, the shock positions are larger than in Cas A by a factor of 1.3–1.5.

The CSM around the WR progenitors (30 and 33 M$_\odot$ models) is strongly affected by the high mechanical luminosity of the WR winds, which completely clears out the surrounding medium up to the termination shock at the inner edge of WR bubble. In the case of the 33 M$_\odot$, the stellar wind interacts directly with the clumps of the broken WR shell producing bow shocks. In both cases, the strong WR winds clear out the CSM at least up to 10pc, so the pre-SN circumstellar medium inside of 10 pc only consists of a low-density, fast wind, with $\rho \propto r^{-2}$ profile. The SN ejecta interacts with this medium at an initial velocity of 9,500 km s$^{-1}$. At 334 yr, the SN forward shock is at 3.1 pc, with ∼ 9,000 km s$^{-1}$, and the reverse shock front is at 1.5 pc, with 5,500 km s$^{-1}$.

### 5.5. Comparison with hydrodynamical models. QSF locations and shocked mass

QSFs in Cas A are distributed around the remnant center with typical radii of ∼ 1.9 - 2 pc. Their enhanced density was one of the reasons to argue in favor of a dense circumstellar shell (Chevalier & Liang 1989). The N enrichment in QSFs seems to be the result of CNO-burning in combination with mass loss from the progenitor. Based on the high N/H ratios, several authors have proposed WN and RSG stars as progenitors, where the QSFs represent clumps ejected from the N-enriched stellar surface with strong density contrasts (∼ $10^2 - 10^3$) compared with a smooth wind (Chevalier & Oishi 2003).

QSFs seem to lie in a circumstellar shell shocked by the SN forward shock; however, our SN precursors develop circumstellar piled-up shells much farther out than 2 pc. Even the nearest computed shell has a larger radius than the forward shock radius of Cas A. In our simulations, the blast wave hits the RSG shell at 830 yr after the explosion, the LBSG shell at 1,100 yr, and the WR shell at 1,670 yr, a much longer time than the age of Cas A.

To reproduce the QSF positions of Cas A, the inner shells in our models must be closer to the star (Schure et al. 2008). Smaller shells could be formed in different ways, for example, if the duration of the latest evolutionary stage (LBSG or WR) is shorter. Also, circumstellar shells could be smaller if their expansion were slower. The velocity V$_s$ of an energy, wind-driven shell expanding in a previous wind is given for example by (Chevalier & Imamura 1983):

$$V_s = \left[\frac{\dot{M}_2 V_2^2 V_1}{3\dot{M}_1}\right]^{1/3} \quad (1)$$

where subindex 1 refers to the slow wind and subindex 2 to the fast wind. If the RSG mass-loss rates were higher and/or LBSG/WR mass loss rates were lower, then the resulted swept-up shells would be slower and smaller. Increased RSG mass loss rates have recently been suggested by Schroder & Cuntz (2005), while lower mass loss rates in WR stars due to clumpiness have been taken into account recently by Hamann & Koesterke (1998), and by Eldridge & Tout (2004). Modifying the mass loss rates, however, would affect the entire stellar evolution so new computations would be required.

Most of the models for the emission of Cas A establish that a fraction of the pre-SN mass loss becomes observable when hit by the blast wave. For example, the x-ray emission in Cas A comes from high-temperature material heated by both SN shocks. The total x-ray emitting mass in Cas A is ∼ 14 M$_\odot$, being approximately 2–4 M$_\odot$ from the SN shocked ejecta and ∼ 8–9 M$_\odot$ from the swept-up mass (Vink et al. 1996, Willingale et al. 2002). Tsunemi et al.(1986), however, found a lower swept-up mass of ∼ 2.4 M$_\odot$, using a non-equilibrium, ionization plasma hypothesis. When we compare our results with the shocked mass in Cas A, we find that only the RSG progenitor is able to provide enough shocked mass due to the dense RSG wind. The LBSG and WR SN progenitors have extended zones of free streaming wind with very low masses, which are insufficient to explain the large amount of emitting mass in Cas A.

## 6. Summary and conclusions

1. The RSG progenitor (23 M$_\odot$ model) agrees quite well with the shock fronts positions and shock fronts ratio observed in Cas A. Due to the dense RSG wind as pre-SN, this kind of progenitor is the one able to explain the large amount of swept-up mass observed in the remnant. The RSG progenitor, however, fails to produce the high N/H and He/H ratios observed in QSFs, if we accept the scenario where QSFs were ejected from the stellar surface as a smooth wind or blobs. The H-rich envelope of the RSG also fails to produce the NKs. The large amount of ejecta mass (11.6 M$_\odot$) during

   the SN is not able to explain the observed ejecta mass (2–4 $M_\odot$) in Cas A, which is ~ 60 % lower.
2. The LBSG progenitors (28 and 29 $M_\odot$ models) are the best candidates for producing NKs as a result of N condensations from their low mass and H-poor envelopes. These stars, however, do not develop QSF abundances in their surfaces, although the He/H and N/H ratios are higher than those from the RSG progenitor. The SN ejecta masses are higher than the observational values by a factor of 2.
3. The first WR progenitor (30 $M_\odot$ model) loses its H envelope but contains enough N onto the surface to produce features similar to NKs. The final stellar mass is low (5.85 $M_\odot$) and the SN ejecta mass (4.40 $M_\odot$) agrees with what is observed in Cas A. Additionally, the 30 $M_\odot$ model shows QSFs ratios on it surface at the beginning of WN-H poor stage.
4. The 33 $M_\odot$ model has the advantages of a WR progenitor, i.e., surface QSF abundances and low SN ejecta mass; however, the WR loses all the N on its surface to produce NKs. The large WR mass-loss rate peels off the star and exposes the core processed by $3\alpha$ reactions.
5. The QSFs in Cas A are located at an average radius of ~ 1.5 pc. None of our models give the correct position because the RSG, LBSG, or WR shells have larger radii. To reproduce the QSF positions and to increase the amount of shocked, swept-up mass as observed in Cas A, the inner shells of our LBSG/WR progenitors must be closer to the star. This can be achieved by higher RSG mass loss rates (Schroeder and Cuntz 2005) and/or by lower LBSG/WR mass loss rates (Hamann and Korestke 1998).
6. There are many uncertainties that should be investigated in future papers, like the mass-loss rate in the various regions of the HR diagram, or the overshooting, which alters the location of the star in the HR diagram and hence affects the mass-loss rate, and also the inclusion of stellar rotation, which produces important changes in both the evolution and the mass-loss rate.

*Acknowledgements.* We would like to thank our anonymous referee for the valuable comments, which have benefited the paper considerably. We would like also to thank Jacco Vink, Klara Schure, Allard Jan van Marle, and Robert Fesen for fruitful discussions. As usual, we also thank Michael L. Norman and the Laboratory for Computational Astrophysics for the use of ZEUS-3D. The computations were performed at the Instituto de Astronomía, Universidad Nacional Autónoma de México. This work has been partially supported by grants from DGAPA-UNAM (IN130698, IN117799 & IN114199) and CONACyT (32214-E).